\documentclass[doublecol]{epl2}
\title{Charge screening and carrier transport in AA-stacked bilayer graphene: tuning via a perpendicular electric field}
\shorttitle{Charge screening and carrier transport in AA-stacked
bilayer graphene}
\author{Yawar Mohammadi}
\shortauthor{Yawar Mohammadi}

\institute{Young Researchers and Elite Club, Kermanshah Branch,
Islamic Azad University, Kermanshah, Iran}
\pacs{77.22.-d}{Dielectric properties of solids and liquids}
\pacs{72.10.-d}{Theory of electronic transport; scattering
mechanisms} \pacs{72.80.Vp}{Electronic transport in graphene}

\abstract{The static dielectric function in AA-stacked bilayer
graphene (BLG), subjected to an electric field applied
perpendicular to layers, is calculated analytically within the
random phase approximation (RPA). This result is used to calculate
the screened Coulomb interaction and the electrical conductivity.
The screened Coulomb interaction, which here can be tuned by the
perpendicular electric field, shows a power-law decay as
$1/(\gamma^{2}+V^2)$ at long-distance limit where $V$ and $\gamma$
are the electrical potential and the inter-layer hopping energy
respectively, indicating that the Coulomb interaction is
suppressed at high perpendicular electric fields. Furthermore, Our
results for the effect of the short-range and the long-range
(Coulomb) scattering on the electrical conductivity show that the
short-range scattering yields a constant electrical conductivity
which is not affected by the perpendicular electric filed. While
the electrical conductivity limited by the Coulomb scattering is
enhanced by the perpendicular electric field and increases
linearly in $V^2$ at small $V$ with a finite value at $V=0$,
indicating that we can tune the electrical conductivity in
AA-stacked BLG by applying a perpendicular electric field.}
\begin{document}
\maketitle

\section{Introduction}

Single layer graphene (SLG), an isolated layer of graphite, shows
many interesting properties~\cite{b.1,b.2} originated from its
chiral linear low energy spectrum dominated by a massless
Dirac-like equation. Stacking order of graphene layers can change
dramatically(or greatly) these properties leading, for example, to
the gapless parabolic spectrum for AB-stacked BLG showing
properties~\cite{b.3,b.4} which are different from those in
graphene. Recently a new stable order of few-layer graphene,
few-layer graphene with AA stacking order, has been observed in
experimental researches~\cite{b.5,b.6}. In this staking order of
graphene layers, each sublattice in a top layer is located
directly above the same one in the bottom layer. They have a
special band structure composed of SLG-like band structures with
different doping which depend on the number of layer and the
inter-layer hopping energy~\cite{b.6,b.7,b.8}. Recently,
AA-stacked BLG has been studied in theoretically, leading to
discovery of some interesting
properties~\cite{b.9,b.10,b.11,b.12,b.13,b.14} which mainly
originate from its special low energy band structure.

One of the most fundamental physical quantities is static
polarization function. Knowing this quantity is essential to study
many fundamental properties, e.g., the RKKY interaction between
magnetic adatoms, Kohn anomaly in phonon dispersion and the
carrier transport through screened coulomb interaction by charged
impurities. The static polarization function of
SLG~\cite{b.15,b.16,b.17,b.18,b.19,b.20,b.21} and AB-stacked
BLG~\cite{b.22,b.23,b.24}, in recent years, have been studied
extensively. The main result of these works is the vanishing
(enhanced) $2k_{F}$ backscattering in SLG (BLG) ($k_{F}$ is the
Fermi wave vector.), which plays a key role in determining low
density and low temperature carrier transport, resulting in
different features for SLG and BLG. In a recent
research~\cite{b.25}, the AA-stacked BLG static polarization has
been calculated analytically. One of aims of this paper is to
calculate analytically the AA-stacked BLG static polarization in
the presence of an electric field applied perpendicular to layers,
which can be use to tune the properties of AA-stacked BLG.

The other purpose of our work is to calculate the carrier
transport in AA-stacked BLG and to investigate effects of the
perpendicular electric filed on it. The carrier transport in
SLG~\cite{b.1,b.15,b.26,b.27,b.28,b.29,b.30} and AB-stacked
BLG~\cite{b.3,b.31,b.32,b.33,b.34} is controlled mainly by two
scattering mechanisms, (i) the long-range Coulomb scattering by
random chared impurities located in the substrate near the
graphene layers and (ii) the the short-range disorder scattering
coming, for example, from the zero range point defects, or
resonant short-range scattering, or other mechanisms. The first is
often dominant in controlling the carrier transport in SLG,
leading to linear dependence of its conductivity on carrier
density beside weak temperature dependence of its conductivity.
While the later becomes important in high mobility samples of SLG.
On the other hand, both scattering mechanisms contribute
significantly to determine the carrier transport in AB-stacked
BLG, leading to the strong insulating temperature dependence of
AB-stacked BLG conductivity beside the linear dependence of
AB-stacked BLG conductivity on carrier density at high carrier
densities. Moreover, AA-stacked BLG lattice has some similarities
with both SLG and AB-stacked BLG. The band structure of AA-stacked
BLG is composed of two SLG-like spectrum. In addition, the
Thomas-Fermi screening wave vector in AA-stacked BLG, similar to
AB-stacked BLG and in contrast to SLG, is constant and independent
of chemical potential. This can lead to different screened Coulomb
impurity potential with respect to that in SLG. Due to these
features, it is reasonable to take into account both short- and
long-range scattering as key scattering mechanisms in controlling
the carrier transport, to study the the carrier transport in
AA-stacked BLG. This is what we want to do in this paper.

This paper is organized as follows. In the section II, we
introduce the tight-binding Hamiltonian describing the low energy
quasiparticle excitation in AA-stacked BLG, subjected to an
electric field applied perpendicular to layers, and obtain
corresponding eigenvalues and eigenfunctions. Section III is
devoted to obtain an analytical relation for the static
polarization function of AA-stacked BLG and we use this result to
consider the effect of the perpendicular electric filed on the
Coulomb interaction in AA-stacked BLG. In section IV, we consider
the electrical conductivity, limited by short- and long-range
scattering, and show how one can tune the electrical conductivity
in AA-stacked BLG by making use of a perpendicular electric field.
Finally, we end our paper by summary and conclusions in section V.

%\section{Model Hamiltonian and Calculations}
\section{Model Hamiltonian}

\begin{figure}
\includegraphics[width=8.75cm,angle=0]{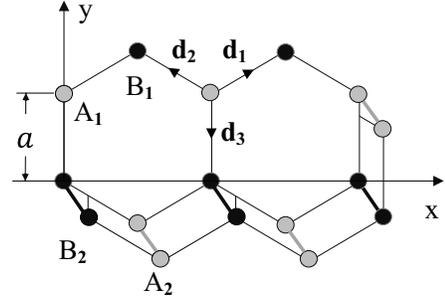}
\caption{A segment of AA-stacked BLG lattice structure.
$\mathbf{d}_{1}$, $\mathbf{d}_{2}$ and $\mathbf{d}_{3}$ are three
vectors that are drown from a sub-site  to its nearest neighbors.}
\label{fig:01}
 \end{figure}

\begin{figure}
\includegraphics[width=8.5cm,angle=0]{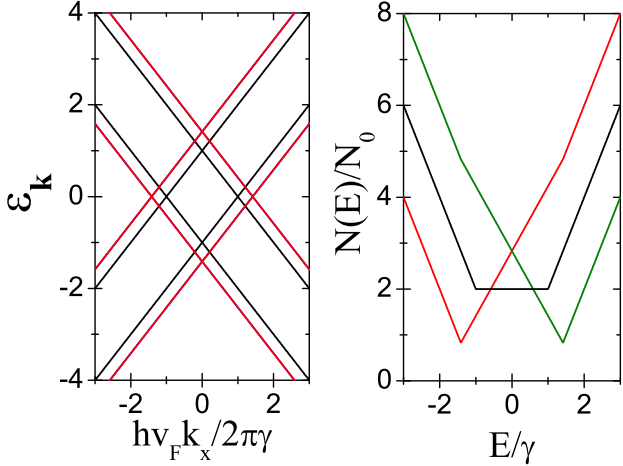}\caption{Left panel
shows the low-energy band structure of biased AA-stacked BLG for
three values of the bias voltage $V=0.0$ (solid line), $V=-\gamma$
(dashed line) and $V=+\gamma$ (doted-dashed line). Right panel
shows the DOS on a sublattice of top layer for these values of the
applied electrical potential with
$N_{0}=4\gamma/\hbar^{2}v_{F}^{2}$ being the DOS at $E=0$.}
\label{fig:02}
\end{figure}

In an AA-stacked BLG lattice which is composed of two SLG, each
sublattice in the top layer is located directly above the same one
in the bottom layer. The unit cell of an AA-stacked BLG consists
of 4 inequivalent Carbon atoms, two atoms for every layer (fig.
\ref{fig:01}). Moreover, the presence of the bias voltage creates
a potential $+V$ in the top layer and $-V$ in the bottom layer.
Thus its Hamiltonian, in the nearest-neighbor tight-binding
approximation, is given by
\begin{eqnarray}
H=\sum_{\mathbf{q}}\hat{\Psi}^{\dag}_{\mathbf{q}}\widehat{H}_{\mathbf{q}}
\hat{\Psi}_{\mathbf{q}}, \label{eq:01}
\end{eqnarray}
where
\begin{eqnarray}
\widehat{H}_{\mathbf{q}} = \left(
\begin{array}{cccccc}
V &   \phi^{\ast}(\mathbf{q})   & \gamma &  0 \\
\phi(\mathbf{q})   & V & 0 & \gamma  \\
 \gamma & 0 & -V &    \phi^{\ast}(\mathbf{q}) \\
 0 & \gamma &   \phi(\mathbf{q})   & -V \\
\end{array}
\right),\label{eq:02}
\end{eqnarray}
and
$\hat{\Psi}^{\dag}_{\mathbf{q}}=(a^{\dag}_{1\mathbf{q}},b^{\dag}_{1\mathbf{q}}
,a^{\dag}_{2\mathbf{q}},b^{\dag}_{2\mathbf{q}})$.
$a^{\dag}_{n\mathbf{q}}$ ($b^{\dag}_{n\mathbf{q}}$) are creation
operators of an electron with momentum $\mathbf{q}$ at A(B)
sublattice in nth-layer. $\gamma$ is the inter-layer hopping
energy and
$\phi(\mathbf{q})=-t\sum_{i=1}^{^{3}}e^{i\mathbf{q}.\mathbf{d}_{i}}$
where $\mathbf{d}_{1}=(a\sqrt{3}/2,a/2)$,
$\mathbf{d}_{2}=(-a\sqrt{3}/2,a/2)$ and $\mathbf{d}_{3}=(0,-a)$
are the nearest neighbor vectors (fig. \ref{fig:01}) with $a$
being the shortest Carbon-Carbon distance.
To obtain the Hamiltonian dominates the low-energy excitations,
which occur near Dirac points ($\mathbf{K}$ and $\mathbf{K}^{'}$),
we must expand $\phi(\mathbf{q})(\phi^{\ast}(\mathbf{q}))$ for
$|\mathbf{k} |\ll|\mathbf{K}|$ around Dirac points where
$\mathbf{q}=\mathbf{k}+\mathbf{K}$. If we expand the Hamiltonian
around $\mathbf{K}=(2\pi/3\sqrt{3}a,2\pi/3a)$ point, we have
$\phi(\mathbf{q})=\hbar v_{F}k_{+}(\phi^{\ast}(\mathbf{q})=\hbar
v_{F}k_{-})$ where $k_{\pm}=k_{x}\pm ik_{y}$ and
$v_{F}=3ta/2\hbar\simeq 9\times 10^{5}m s^{-1}$ is Fermi velocity.
Our results for the corresponding low-energy eigenvalues and
eigenstates are given by
\begin{eqnarray}
\varepsilon_{\mathbf{k}}^{s\lambda}=s\gamma^{'}+\lambda
\hbar v_{F}|\mathbf{k}|,\nonumber \\
\psi^{s\lambda}_{\mathbf{k}}=\frac{\gamma}{2\sqrt{\gamma^{'}(\gamma^{'}-V)}}
\left(
\begin{array}{c}
1\\
\lambda e^{-i\theta_{\mathbf{k}}}\\
s\frac{\gamma^{'}-V}{\gamma}\\
s\lambda\frac{\gamma^{'}-V}{\gamma}e^{-i\theta_{\mathbf{k}}}\\
\end{array}
\right),\label{eq:03}
\end{eqnarray}
where $s=\pm$, $\lambda=\pm$ and
$\gamma^{'}=\sqrt{\gamma^{2}+V^{2}}$. Here
$|\mathbf{k}|=\sqrt{k_{x}^{2}+k_{y}^{2}}$ is the amount of the
two-dimensional momentum measured from Dirac points and
$\theta_{\mathbf{k}}=tan^{-1}(k_{y}/k_{x})$. Notice, the low
energy band structure of AA-stacked BLG is a composition of two
electron-doped and hole-doped SLG-like band
structures~\cite{b.6,b.7,b.8}, which for some properties behave
like decoupled bands leading to many attractive properties not
been observed in the other graphene-based materials so far
~\cite{b.9,b.10,b.11,b.12,b.13,b.14}. The low energy density of
states of AA-stacked BLG is
\begin{eqnarray}
D(E)=g\frac{|E-\gamma^{'}|+|E+\gamma^{'}|}{2\pi \hbar^{2}
v_{F}^{2}},\label{eq:04}
\end{eqnarray}
where multiple $g=4$ is due to the spin and valley degeneracies.
The low energy band structure and the density of states of
AA-stacked BLG, for different values of the bias voltage, have
been shown in fig. \ref{fig:02}.

\section{Static polarization function and charge-carrier screening}
\label{sec:2}

The static dielectric function in the low energy excitation limit,
where inter-layer and intra-layer Coulomb interactions are
approximately equal, and within RPA is given by
\begin{eqnarray}
\epsilon(q)=1+\frac{2\pi e^{2}}{\kappa q}\Pi(q),\label{eq:05}
\end{eqnarray}
where $\Pi(q)$ is the static polarization function and $\kappa$ is
the background dielectric constant.
 The static polarization function of
AA-stacked BLG is given by
\begin{eqnarray}
\Pi(q)=-\frac{g}{2A} \sum_{\mathbf{k}\lambda\lambda^{'}ss^{'}
}\frac{n_{\mathbf{k}}^{s\lambda}-n_{\mathbf{k}^{'}}^{s^{'}\lambda^{'}}}
{\varepsilon_{\mathbf{k}}^{s\lambda}-\varepsilon_{\mathbf{k}^{'}}^{s^{'}\lambda^{'}}}
F^{ss^{'}\lambda\lambda^{'}}_{\mathbf{k}\mathbf{k}^{'}},\label{eq:06}
\end{eqnarray}
where $\mathbf{k}^{'}=\mathbf{k}+\mathbf{q}$, $\Delta
\theta_{\mathbf{k}^{'},\mathbf{k}}=\theta_{\mathbf{k}^{'}}-\theta_{\mathbf{k}}$
and $F^{ss^{'}\lambda\lambda^{'}}_{\mathbf{k}\mathbf{k}^{'}}$ is
the overlap of electron and hole wave-function,
$F^{ss^{'}\lambda\lambda^{'}}_{\mathbf{k}\mathbf{k}^{'}}=|\langle
\psi^{s^{'}\lambda^{'}}_{\mathbf{k}^{'}}
|e^{i(\mathbf{k}^{'}-\mathbf{k}).\mathbf{r}}|
\psi^{s\lambda}_{\mathbf{k}} \rangle|^{2}$, which becomes
$(1+\lambda\lambda^{'}\cos
\Delta\theta_{\mathbf{k}^{'},\mathbf{k}})/2$ if $s=s^{'}$ and is
zero if $s\neq s^{'}$. This indicates that only transitions
between electron and hole bands with sam $s$(or $s^{'}$)
band-index contribute to the polarization function of AA-stacked
BLG. Here,
$n_{\mathbf{k}}^{s\lambda}=1/(1+exp[(\varepsilon_{\mathbf{k}}^{s\lambda}-\mu)/k_{B}T])$
is the Fermi-Dirac distribution function with $k_{B}$ being
Boltzmann constant.

In this paper, we only calculate the polarization function of the
AA-stacked BLG in undoped regime at zero temperature. In this case
the static polarization function of AA-stacked BLG can be written
as sum of two SLG static polarization function with
$\mu=\gamma^{'}$ and $\mu=-\gamma^{'}$ which, as it has been shown
before~\cite{b.8}, are equal. Therefore, the static polarization
function of undoped AA-stacked BLG is equal to that of a doped SLG
with $\mu=\gamma^{'}$ which can be written as
\begin{eqnarray}
\Pi(q)&=&+\frac{g}{A}\sum_{\mathbf{k}}\frac{1-\cos
\Delta\theta_{\mathbf{k},\mathbf{k+q}}} {\hbar
v_{F}(|\mathbf{k}|+|\mathbf{k+q}|)} \nonumber
\\
&&-\frac{g}{A}\sum_{\mathbf{k}\lambda}\frac{1-\lambda\cos
\Delta\theta_{\mathbf{k},\mathbf{k+q}}} {\hbar
v_{F}(|\mathbf{k}|+\lambda|\mathbf{k+q}|)}
\theta(\gamma^{'}-v_{F}|\mathbf{k}|),\label{eq:07}
\end{eqnarray}
where the first term, which is equal to the static polarization of
the undoped SLG, is $gq/16\hbar v_{F}$. The second term can be
easily calculated similar to what done in Ref.~\cite{b.8,b.15}.
Hence, we have $\Pi(q)=\frac{g\gamma^{'}}{2\pi \hbar^{2}
v_{F}^{2}}$, for $q\leq 2\gamma^{'}/\hbar v_{F}$, and
\begin{eqnarray}
\Pi(q)&=& \frac{g
q}{16 \hbar v_{F}}+\frac{g\gamma^{'}}{2\pi \hbar^{2} v_{F}^{2}}(1-\nonumber \\
&& \frac{1}{2}\sqrt{1-(\frac{2\gamma^{'}}{\hbar v_{F}q})^{2}}
-\frac{\hbar v_{F}q}{4\gamma^{'}}\sin^{-1}\frac{2\gamma^{'}}{\hbar
v_{F}q})
 ,\label{eq:08}
\end{eqnarray}
for $q> 2\gamma^{'}/\hbar v_{F}$. Notice that the static
polarization function of AA-stacked BLG, for $q\leq
2\gamma^{'}/\hbar v_{F}$, ( similar to that in doped SLG and
ordinary 2DEG) is a constant metallic-like polarization, even in
the absence of doping. This constant polarization, in zero limit
of the perpendicular electric field, is only depend on the
inter-layer hopping energy and on the Fermi velocity $v_{F}$.
Moreover, it can be tuned by a perpendicular electric field. For
$q> 2\gamma^{'}/\hbar v_{F}$, AA-stacked BLG, similar to SLG, has
a insulating-like polarization which increases linearly in $q$.
Furthermore, the value of a momentum, at which a crossover from
metallic to insulating screening takes place ($q=2\gamma^{'}/\hbar
v_{F}$), can be tuned electrically and this allow us to suppress
the insulating screening effects via a perpendicular electric
filed.

The static screening, in the long wave-length limit, is given by
$\epsilon(q)\approx1+q_{TF}/q$ where $q_{TF}$ is the Thomas-Fermi
wave-vector. For the biased AA-stacked BLG, Thomas-Fermi
wave-vector is $q_{TF}=2\pi
e^{2}D(0)/\kappa=ge^{2}\sqrt{\gamma^{2}+V^{2}}/2\kappa \hbar^{2}
v_{F}^{2}$ which, similar to that in ordinary 2DEG and in contrast
to that in SLG, is independent of carrier concentration.

Moreover, the electrical field dependence of the static dielectric
function allow us to tune charge screening in AA-stacked via an
electric field applied perpendicular to layers and manipulate some
attractive properties of AA-stacked BLG~\cite{b.11} arising from
Coulomb interaction of electrons. This feature can be seen,
explicitly, in electrical potential dependence of the
long-distance behavior of Coulomb interaction. The long-distance
behavior of long-range Coulomb interaction consist of two parts, a
non-oscillatory term coming from long wave-length behavior of the
static polarization (Thomas-Fermi approximation) and a
Friedel-oscillation part arising from a discontinuity occurring in
the second derivative of the static polarization. The
non-oscillatory part, which can obtained by making use of
Thomas-Fermi dielectric function, is given by
\begin{eqnarray}
\phi(r)=\frac{Ze^{2}}{\kappa r}-\frac{\pi Ze^{2}q_{TF}
}{2\kappa}[H_{0}(q_{TF}r)-Y_{0}(q_{TF}r)],\label{eq:09}
\end{eqnarray}
with $H_{0}$ and $Y_{0}$ being the Struve and the Bessel functions
of the second kind. The asymptotic form of this term at large
distance is given by $Ze^{2}q_{TF}/[\kappa(q_{TF}r)^{3}]$ where
$q_{TF}=ge^{2}\sqrt{\gamma^{2}+V^{2}}/2\kappa \hbar^{2}
v_{F}^{2}$. It is evident that by increasing $V$ this part
decreases as $1/(\gamma^{2}+V^{2})$ leading to a suppressed
interaction at high electric field.

The Friedel-oscillation part, which originates from a
discontinuity occurring in the second derivative of the static
polarization, can be calculated by making use of a theorem of
Lighthill~\cite{b.35}. To obtain these terms we must use the
asymptotic form of the the Bessel function. Therefore, we have
\begin{eqnarray}
\phi(r)\simeq Ze^{2} \frac{\sqrt{k_{F}^{'}}}{\sqrt{\pi r}}
\int_{0}^{\infty}
\frac{\cos(k_{F}^{'}rx)+\sin(k_{F}^{'}rx)}{x+\frac{2\pi
e^{2}}{k_{F}^{'}}\Pi(x)}\sqrt{x}dx,\label{eq:10}
\end{eqnarray}
where $k_{F}^{'}=\sqrt{\gamma^{2}+V^{2}}/\hbar v_{F}$,
$x=q/k_{F}^{'}$. This integral can be easily calculated~\cite{b.8}
which becomes
\begin{eqnarray}
\phi(r)\simeq -\frac{3Ze^{2}}{4\kappa} \frac{\alpha
k_{F}^{'}}{(1+\pi \alpha)^{2}}
\frac{\cos(2k_{F}^{'}r)}{(k_{F}^{'}r)^{3}},\label{eq:11}
\end{eqnarray}
where $\alpha=e^{2}/\kappa \hbar v_{F}$.  Friedel-oscillation
part, similar to non-oscillatory part, depend on electrical
potential as $1/(\gamma^{2}+V^{2})$, but with an extra oscillatory
coefficient, $\cos(2r\sqrt{\gamma^{2}+V^{2}}/\hbar v_{F})$ whose
periodicity can be tuned by the electric field.

\section{Carrier transport}
\label{sec:4}

We use Boltzmann equation to calculate the carrier transport in
AA-stacked BLG, motivated by this fact that the theoretical
results obtained from this equation for the carrier transport in
SLG~\cite{b.1,b.15,b.26,b.29,b.30} and AB-stacked
BLG~\cite{b.3,b.32,b.33} are in good agreement with the reported
experimental results. It is logical to suppose that the charge
carrier in AA-stacked BLG, even in absence of doping, behave like
a homogenous electron gas. This is due to the large density of
state at Fermi level, the average carrier density is always larger
than the fluctuations in carrier density and this prevents from
formation of electron-hole puddle structures induced by the
charged impurities which is observed in SLG and AB-stacked BLG
close to the charge neural point. Moreover, since the low energy
bands with different s index band in AA-stacked BLG behave like
decoupled bands, the electrical conductivity of AA-stacked BLG can
be written as sum of two terms for SLG electrical conductivity
with $E_{F}=-\gamma^{'}$ and $E_{F}=+\gamma^{'}$. The electrical
conductivity in a homogenous electron gas of massless chiral Dirac
charged carriers is given by
\begin{eqnarray}
\sigma=g\frac{e^{2}}{4\pi \hbar^{2}}\int d\epsilon_{\mathbf{k}}
\tau(\epsilon_{\mathbf{k}})\epsilon_{\mathbf{k}}(-\frac{\partial
f(\epsilon_{\mathbf{k}}) }{\partial
\epsilon_{\mathbf{k}}})\label{eq:12}
\end{eqnarray}
where $\epsilon_{\mathbf{k}}=\hbar v_{f}|\mathbf{k}|$,
$f(\epsilon_{\mathbf{k}})=[1+exp((\epsilon_{\mathbf{k}}-\mu)/k_{B}T)]^{-1}$
is Fermi distribution function with $\mu$ being chemical potential
and $\tau(\varepsilon_{\mathbf{k}})$ is the scattering time given
by
\begin{eqnarray}
\frac{1}{\tau(\epsilon_{\mathbf{k}})}=&&\frac{\pi}{\hbar}\int
\frac{d\mathbf{k}^{'}}{(2\pi)^{2}}
\frac{n_{s}|v_{s}(q)|^{2}+n_{l}|v_{l}(q)|^{2}}{(\epsilon(q))^{2}}\nonumber
\\ &\times& (1-\cos^{2}\theta_{\mathbf{k}\mathbf{k}^{'}})
\delta(\epsilon_{\mathbf{k}}-\epsilon_{\mathbf{k}^{'}}),\label{eq:13}
\end{eqnarray}
where $v_{l}(q)$($v_{s}(q)$) is the matrix elements of the
long-(short-)range scattering potential between an electron and an
charged impurity (a point defect) and  $n_{l}$($n_{s}$) is the
corresponding impurity density. The long-range Coulomb interaction
is given by $v_{l}(q)=2\pi e^{2}e^{-qd}/\kappa q$ with $d$ being
the average distance of charged impurity from AA-stacked BLG. The
short-range interaction is $v_{s}(q)=v_{0}=const$. In this paper
we only consider zero temperature case. Therefore we have
$\sigma=\frac{e^{2}}{h}\frac{2E_{F} \tau(E_{F})}{\hbar}$ where
$\tau(E_{F})$ is the scattering time at zero temperature .

\begin{figure}
\includegraphics[width=8.5cm,angle=0]{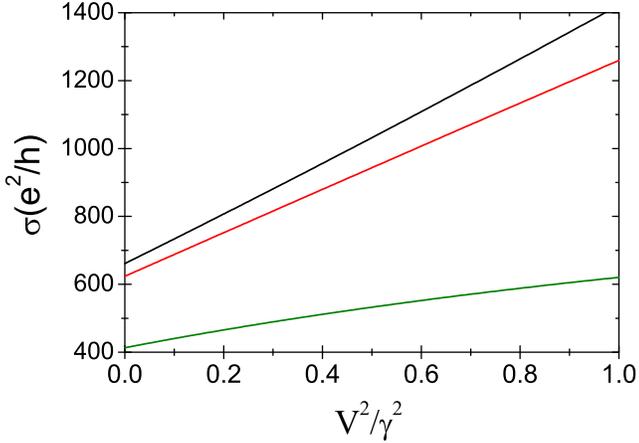}
\caption{Electrical Conductivity in AA-stacked BLG as a function
of $V^{2}$ calculated numerically from Eqs. (\ref{eq:14}) and
(\ref{eq:15}) for $n_{l}=10^{12} cm^{-2}$, $v_{0}=1 keV$ and $d=4$
. Black, red and green lines are correspond to $n_{s}/n_{l}=0$,
0.5 and 0.05 respectively.} \label{fig:03}
\end{figure}

It is interesting to compare the role of the short-range and
Coulomb scattering in controlling the electrical conductivity in
AA-stacked and also to see how the perpendicular electric filed
affects on each contribution. We can write the electrical
conductivity as
$\frac{1}{\sigma}=\frac{1}{\sigma_{s}}+\frac{1}{\sigma_{l}}=\frac{h}{e^{2}}(\Lambda_{s}+\Lambda_{l})$,
where $\sigma_{s}^{-1}$ and $\sigma_{l}^{-1}$ are the electrical
resistivity arising from short-rang and Coulomb scattering
respectively, and $\Lambda_{s}$ and $\Lambda_{l}$ are given by
\begin{eqnarray}
\Lambda_{s}=\frac{2n_{s}v_{0}^{2}}{\pi\hbar^{2}
v_{F}^{2}}\int_{0}^{1}d\lambda
\frac{\lambda^{4}\sqrt{1-\lambda^{2}}}{(\lambda+4r_{s})^{2}}
\label{eq:14}
\end{eqnarray}
and
\begin{eqnarray}
\Lambda_{l}=\frac{2 n_{l}r_{s}^{2}\hbar^{2}
v_{F}^{2}}{\gamma^{2}+V^{2}}\int_{0}^{1}d\lambda
\frac{\lambda^{2}\sqrt{1-\lambda^{2}}}{(\lambda+4r_{s})^{2}}e^{-\frac{4\sqrt{\gamma^{2}+V^{2}}}{3t}\frac{d}{a}\lambda}
\label{eq:15}
\end{eqnarray}
with $\lambda=\hbar v_{F}q/2\sqrt{\gamma^{2}+V^{2}}$ and
$r_{s}=e^{2}/\hbar v_{F}\kappa$ being the dimensionless
Wigner-Seitz radius in graphene which is a constant. It is evident
that, due to the nonzero density of state at Fermi energy level,
even in the absence of doping and at $V=0$, AA-stacked BLG shows a
finite electrical conductivity. Moreover, Eq. (\ref{eq:14}) shows
that the short-range scattering yields a constant electrical
conductivity which only changes by varying the substrate resulting
in different substrate dielectric constant (and consequently
different $r_{s}$). Another result, which is more interesting, is
that we can enhance $\sigma_{l}$ by applying a perpendicular
electric filed (black line in fig. \ref{fig:03}). This term
increases linearly in $V^{2}$ (black line in fig. \ref{fig:03})
with a nonzero value at $V=0$ which depends on the interlayer
hopping energy and $r_{s}$. This is similar to what has been
reported for SLG~\cite{b.15,b.26,b.27,b.28,b.29,b.30}. This can be
explained by this fact that, due to the special staking order of
AA-stacked BLG, the electrical conductivity of AA-stacked BLG is
equal to that of a doped SLG with finite $k_{F}$ which in the
presence of an electric field applied perpendicular to layers is
$\frac{\sqrt{\gamma^{2}+V^{2}}}{\hbar v_{F}}$. Furthermore, in SLG
the electrical conductivity increases linearly in carrier
concentration or equivalently linearly in $k_{F}^{2}$. So it is
reasonable to obtain a relation for the AA-stacked BLG
conductivity which increases linearly in $V^{2}$ with a finite
value at $V=0$ (Notice that
$e^{-\frac{4\sqrt{\gamma^{2}+V^{2}}}{3t}\frac{d}{a}\lambda}\simeq
0.68$ for $V=\gamma$, $a=1.42 A$ and $d=4 A$ even when
$\lambda=1$). These features provide high potential applicability
for AA-stacked BLG in nanoelectronic devices.

Moreover fig. \ref{fig:03} shows that when $n_{s}/n_{l}\ll 1$ the
Coulomb scattering is dominant and the electrical conductivity
increases linearly in $V^{2}$(black and red lines in fig.
\ref{fig:03}), but for large $n_{s}/n_{l}$ (green line in fig.
\ref{fig:03}) the short-range scattering becomes important,
leading to sub-linear dependence of the conductivity on $V^{2}$ at
large $V$ similar to what has been reported for the conductivity
of high mobility sample of SLG~\cite{b.28,b.29}.

\section{Summary and conclusions}
\label{sec:5}

In summary, we first calculate analytically the static dielectric
function in AA-stacked BLG in the presence of an electric field
applied perpendicular to layers within the random phase
approximation. Then we obtained analytical relations for the
Friedel-oscillation and the non-oscillatory parts of the
long-distance limit of the screened Coulomb interaction, which
shows explicitly their dependence on the applied perpendicular
electric field. This expression revealed that the Coulomb
interaction in AA-stacked BLG is suppressed at the high
perpendicular electric fields. Finally we used the Boltzmann
transport theory to calculate the electrical conductivity in
AA-stacked BLG, examining the effects of Coulomb and short-range
scattering mechanisms. Our results showed that the short-range
scattering, which can arise for example from point defect, yields
a constant electrical conductivity which is independent of the
applied perpendicular electric field and can only change by
varying the substrate and of course by varying the short-range
impur density. On the other hand, we found that the
Coulomb-scattering-limited electrical conductivity can be tuned by
applying a perpendicular electric field, showing a linear
dependence on $V^{2}$ at small perpendicular electric fields.
Moreover we found that when $n_{s}/n_{l}$ increases the electrical
conductivity shows a sub-linear dependence on $V^{2}$ at large
perpendicular electric fields.

%%%%%%%%%%%%%%%%%%%

\begin{thebibliography}{0}
\bibitem{b.1}
  \Name{Castro Neto A. H., Guinea F., Peres N. M. R., Novoselov K. S. \and Geim A. K.}
  \REVIEW{Rev. Mod. Phys.}{81}{2009}{109}.
\bibitem{b.2}
  \Name{Peres N. M. R.}
  \REVIEW{Rev. Mod. Phys.}{82}{2010}{2673}.
%
\bibitem{b.3}
  \Name{Das Sarma S., Adam S., Hwang E. H. \and Rossi E.}
  \REVIEW{Rev. Mod. Phys.}{83}{2011}{407}.
\bibitem{b.4}
  \Name{McCann E., \and Koshino M.}
  \REVIEW{Rep. Prog. Phys.}{76}{2013}{056503}.
%
\bibitem{b.5}
  \Name{Liu Z., Suenaga K., Harris P. J. \and Iijima S.}
  \REVIEW{Phys. Rev. Lett.}{102}{2009}{015501}.
\bibitem{b.6}
  \Name{Borysiuk J., Soltys J. \and Piechota J.}
  \REVIEW{J. Appl. Phys.}{109}{2011}{093523}.
%
\bibitem{b.7}
  \Name{Ando T.}
  \REVIEW{J. Phys.: Conf. Ser.}{302}{2011}{012015}.
\bibitem{b.8}
  \Name{Mohammadi Y. Moradian R. \and Sirzadi Tabar F.}
  \REVIEW{Solid State Commun.}{193}{2014}{1}.
%
\bibitem{b.9}
  \Name{Hsu Y. -F. \and Guo G. -Y.}
  \REVIEW{Phys. Rev. B}{82}{2011}{165404}.
\bibitem{b.10}
  \Name{Tabert C. J. \and Nicol E. J.}
  \REVIEW{Phys. Rev. B}{84}{2012}{075439}.
\bibitem{b.11}
  \Name{Brey L. \and Fertig H. A.}
  \REVIEW{Phys. Rev. B}{87}{2013}{115411}.
\bibitem{b.12}
  \Name{Sboychakov A.O., Rakhmanov A.L., Rozhkov A.V., \and Nori  F.}
  \REVIEW{Phys. Rev. B}{87}{2013}{121401(R)}.
\bibitem{b.13}
  \Name{Sanderson M., Ang Y., \and Zhang C.}
  \REVIEW{Phys. Rev. B}{88}{2013}{245404}.
\bibitem{b.14}
  \Name{Mohammadi Y. \and Moradian R.}
  \REVIEW{Physica B}{442}{2014}{66}.
%
\bibitem{b.15}
  \Name{Ando T.}
  \REVIEW{J. Phys. Soc. Jpn.}{75}{2006}{074716}.
\bibitem{b.16}
  \Name{Katsnelson M. I.}
  \REVIEW{Phys. Rev. B}{74}{2006}{201401}.
\bibitem{b.17}
  \Name{Wunsch B., Stauber T., Sols F. \and Guinea F.}
  \REVIEW{New. J. Phys.}{8}{2006}{318}.
\bibitem{b.18}
  \Name{Hwang E. H. \and Das Sarma S.}
  \REVIEW{Phys. Rev. B}{75}{2007}{205418}.
\bibitem{b.19}
  \Name{Pyatkovskiy P. K., \and Gusynin  V. P.}
  \REVIEW{Phys. Rev. B}{83}{2011}{075422}.
\bibitem{b.20}
  \Name{Sodemann I., \and Fogler M. M.}
  \REVIEW{Phys. Rev. B}{86}{2012}{115408}.
\bibitem{b.21}
  \Name{Scholz A., Stauber T., \and Schliemann J.}
  \REVIEW{Phys. Rev. B}{86}{2012}{195424}.
%
\bibitem{b.22}
  \Name{Hwang E. H. \and Das Sarma S.}
  \REVIEW{Phys. Rev. Lett.}{101}{2008}{156802}.
\bibitem{b.23}
  \Name{Gamayun O. V. }
  \REVIEW{Phys. Rev. B}{84}{2011}{085112}.
\bibitem{b.24}
  \Name{Triola C. \and Rossi E.}
  \REVIEW{Phys. Rev. B}{86}{2012}{161408}.
%
\bibitem{b.25}
  \Name{Roldan R. \and Brey L.}
  \REVIEW{Phys. Rev. B}{88}{2013}{115420}.
%
\bibitem{b.26}
  \Name{Peres N. M. R.}
  \REVIEW{Phys. Mod. Phys.}{82}{2010}{2673}.
\bibitem{b.27}
  \Name{Ziegler K.}
  \REVIEW{Phys. Rev. Lett.}{97}{2006}{266802}.
\bibitem{b.28}
  \Name{Nomura K., \and MacDonald A. H.}
  \REVIEW{Phys. Rev. Lett.}{98}{2007}{076602}.
\bibitem{b.29}
  \Name{Hwang E. H., Adam S. \and Das Sarma S.}
  \REVIEW{Phys. Rev. Lett.}{98}{2007}{186806}.
\bibitem{b.30}
  \Name{Stauber T., Peres N. M. R., \and Guinea F.}
  \REVIEW{Phys. Rev. B 76}{76}{2007}{205423}.
%
\bibitem{b.31}
  \Name{Koshino M. \and Ando T.}
  \REVIEW{Phys. Rev. B}{73}{2006}{245403}.
\bibitem{b.32}
  \Name{Das Sarma M. S., Hwang E. H., \and Rossi E.}
  \REVIEW{Phys. Rev. B}{81}{2010}{161407}.
\bibitem{b.33}
  \Name{Lv M., \and Wan Sh.}
  \REVIEW{PPhys. Rev. B}{81}{2010}{195409}.
\bibitem{b.34}
  \Name{Yuan Sh., De Raedt H., \and Katsnelson M. I.}
  \REVIEW{Phys. Rev. B}{82}{2010}{235409}.
  \bibitem{b.35}
  \Name{Lighthill M. J.}
  \Book{Introduction to Fourier Analysis and Generalised Functions}
%  \Editor{A. Editor}
%  \Vol{9}
  \Publ{Cambridge University Press, Cambridge}
  \Year{1958}
  \Page{52}.
%%%%%%%%%%
%%%%%%%%%%
\end{thebibliography}
\end{document}